\documentstyle[11pt,paspconf,epsf]{article}

\begin{document}
\title{The Galactic Lithium Evolution Revisited}

\author{Donatella Romano\altaffilmark{1}}
\affil{Osservatorio Astronomico di Trieste, 
       via G.B. Tiepolo 11, 34131 Trieste--Italy}
\altaffiltext{1}{SISSA/ISAS, via Beirut 2--4, 34014 Trieste--Italy. 
                 E-mail: romano@sissa.it}

\begin{abstract}
We study the temporal evolution of the $^7$Li abundance in the gaseous 
medium for all the components (halo, disk and bulge) of our Galaxy. We 
adopt nucleosynthesis prescriptions on stellar lithium production based 
upon recent observational results and up-to-date theoretical considerations 
in the framework of a fully consistent chemical evolution model. $^7$Li 
production by AGB stars and SNeII is discussed and the contribution by 
novae is reintroduced. To gain some insight into the stellar factories which 
should be taken into account to reproduce the upper envelope of the 
observational A(Li) vs [Fe/H] diagram for the solar neighborhood, we 
critically analyse Li data from several authors. The contribution from nova 
outbursts seems to be fundamental in order to reproduce the high $^7$Li 
abundances observed in the most metal-rich dwarfs. Finally, we make 
predictions on the $^7$Li content that should be expected in the most 
metal-rich stars of the Galactic bulge. To test similar predictions, $^7$Li 
observations in a statistical significant sample of high metallicity bulge 
stars are absolutely necessary; the advent of \emph{microlensing} should help 
us in reaching this goal.
\end{abstract}

\section{Introduction}

$^7$Li is a key element in quite a lot of astrophysical and cosmological 
problems:
\begin{itemize}
\item it is one of the few elements synthetized during the Big Bang, so its 
primordial abundance (tied to the parameter $\eta$) can in principle be used  
to constrain primordial nucleosynthesis models;
\item there is both theoretical and observational evidence that $^7$Li can be 
produced and/or destroyed in stars, the production and depletion mechanisms 
being so complex and so far from being understood that one may thinks there 
is a cosmic conspiracy that keeps us from drawing any firm picture of the 
$^7$Li abundance evolution in the Galaxy!
\end{itemize}
A key point generally accepted is that the $^7$Li abundance in warm halo 
dwarfs is nearly constant and representative of the pristine $^7$Li abundance 
(Spite \& Spite 1982; Rebolo et al.\,1988; Bonifacio \& Molaro 1997). 
Keeping this as a starting point, chemical evolution models have been 
constructed which explain the rise off the so-called Spite plateau by means of 
stellar $^7$Li production (D'Antona \& Matteucci 1991; Matteucci et al.\,
1995). We will present new chemical evolution results on the temporal 
variation of the $^7$Li abundance in the ISM for both the solar neighborhood 
and the Galactic bulge.

We will start by illustrating the data-sample we have used to constrain the 
chemical evolution model (\S 2); then we will present the chemical evolution 
model itself, focusing mainly on the nucleosynthesis prescriptions used to 
account for $^7$Li production from different stellar sources (\S 3); finally, 
we will show the predicted trends of A(Li) versus [Fe/H] for both the solar 
neighborhood and the Galactic bulge and draw some conclusions (\S 4).

\section{Observational data}

We selected from the literature a great number of $^7$Li measurements in 
stellar atmospheres, with the aim of exploring in particular the A(Li) versus 
[Fe/H] behaviour in the metallicity range -\,1.5 $\le$ [Fe/H] $\le$ -\,0.5 
dex. Our main fonts are:
\begin{itemize}
\item Deliyannis et al.\,(1990)
\item Lambert et al.\,(1991)
\item Pilachowski et al.\,(1993)
\item Pasquini et al.\,(1994)
\item Spite et al.\,(1996)
\end{itemize}
We considered only stars with HIPPARCOS parallax and proper motion, in order 
to give a homogeneous and accurate estimate of their membership to the 
different Galactic kinematical components. To do that, we evaluated u, v and 
w (the usual Galactic space-velocity components with respect to the Local 
Standard of Rest, in a left-handed coordinate system) by means of the 
Johnson \& Soderblom (1987) matrices. With further selection criteria we 
also identified all those stars that we know to have already suffered Li 
depletion (objects with T$_{\mathrm{eff}}$ $<$ 5700 K, see Ryan \& Deliyannis 
1998) or dilution (evolved stars, identified by using theoretical isochrones 
from Bertelli et al.\,1994). Disk stars in our sample, selected on the basis 
of their space-velocity components (see Fig.1 caption for the selection 
criteria) have mostly metallicities greater than -\,1.0 dex and show a large 
scatter in their lithium abundances (Fig.1); all the others show Li depletion 
starting from very low metallicities; moreover, they show no evidence of a 
rise off the plateau values. For stars with multiple $^7$Li detections lying 
on the upper envelope, we take an averaged value for both A(Li) and [Fe/H] 
(see Romano et al.\,1999); we do not take into account the upper limits, 
because they are not significant in order to constrain chemical evolution 
models.

\begin{figure}
\plotone{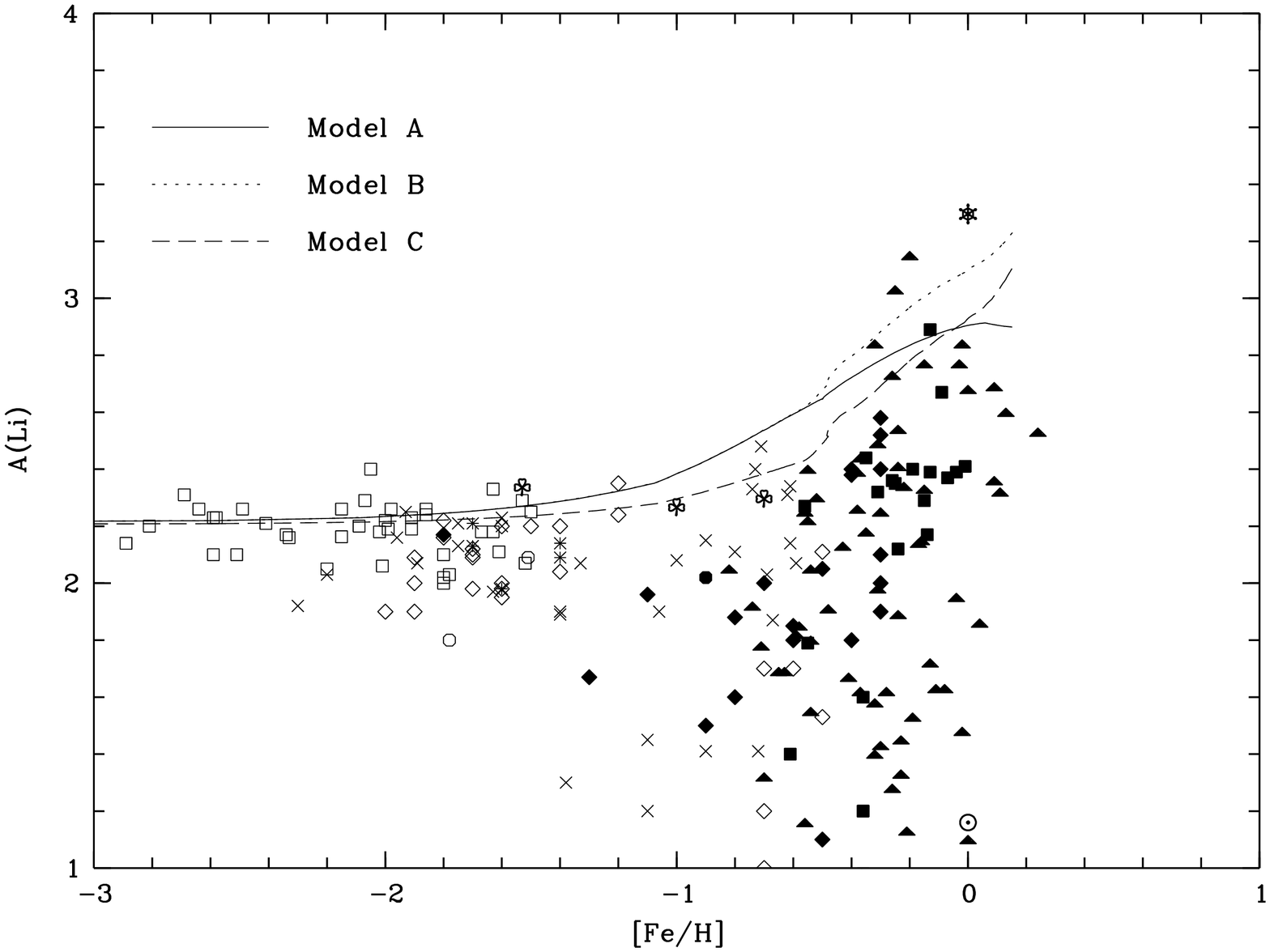}
\caption{ A(Li) vs [Fe/H] diagram for the solar neighborhood. Theoretical 
trends obtained under different nucleosynthesis assumptions are superimposed 
to the observational points. Disk stars (defined as those with v $>$ -\,115 km 
s$\,^{-1}$, (u$^2$\,+\,w$^2$)$\,^{\frac{1}{2}}$ $<$ 150 km s$\,^{-1}$) are 
shown as filled symbols; non disk stars (defined as objects that do not 
satisfy simultaneously the conditions on u, v, w given above) are shown as 
empty symbols. Asterisks: halo stars from Spite et al.\,1996; filled squares: 
Pasquini et al.\,1994; empty squares: halo stars from Bonifacio \& Molaro 
1997; circles: Pilachowski et al.\,1993; triangles: Lambert et al.\,1991; 
lozenges: Deliyannis et al.\,1990. Crosses are stars for which we do not have 
u, v, w determinations; clovers are averaged values for stars on the upper 
envelope for which we have found multiple $^7$Li detections in the literature. 
(Details on data analysis can be found in Romano et al.\,1999.) $^7$Li 
abundances in the Sun (Anders \& Grevesse 1989) and in T-Tauri stars are also 
shown with different symbols.}
\end{figure}

Plotting the target stars in a diagram with rotational velocities around the 
Galactic center as absciss\ae\ and non-rotational velocities as ordinates 
(Fig.2), we can see how, in general, stars with metallicities greater than 
-\,1.0 show disk-like features (although there are some exceptions). From 
Fig.1 and Fig.2 we conclude that the stellar Li content is likely to be 
independent of the star kinematics, although more data are needed to draw firm 
conclusions.

\begin{figure}
\plotone{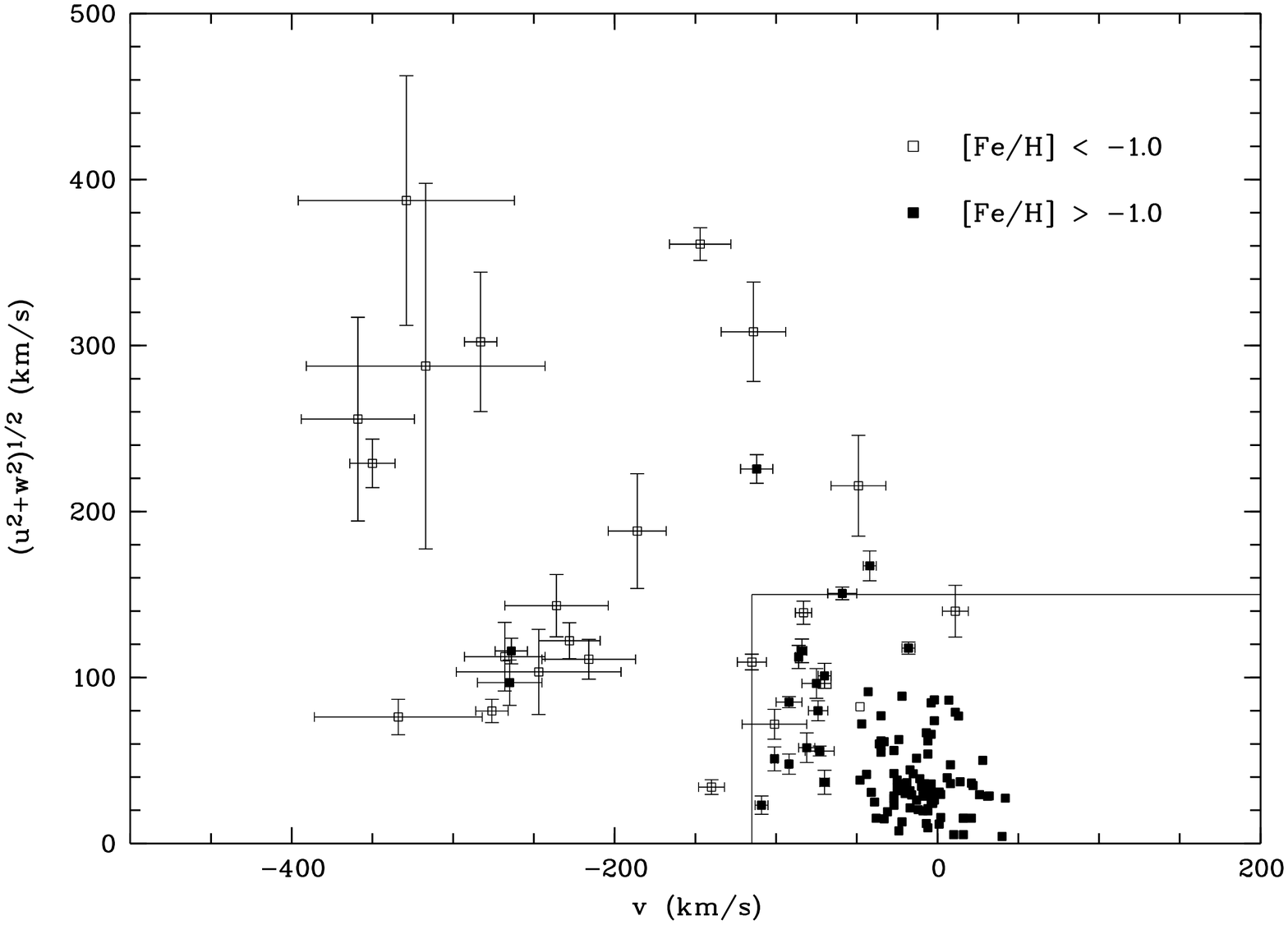}
\caption{ The (u$^2$\,+\,w$^2$)$\,^{\frac{1}{2}}$ vs v diagram for our target 
stars. Two different ranges in metallicity have been considered. The box in 
the low right corner tentatively puts limits on the minimum rotational 
velocity around the Galactic center and on the maximum kinetic energy 
associated to non-rotative motions for a star to be ascribed to the disk 
(either thin or thick).}
\end{figure}

\section{Theoretical models}

To reproduce the upper envelope of the observed diagram Li-metallicity, under 
the hypothesis that it is indicative of the ISM enrichment during the Galactic 
lifetime, we adopt the two-infall chemical evolution model of Chiappini et 
al.\,(1997). It assumes that the Galaxy formed out of two main infall 
episodes. During the first and faster episode, both the halo and the bulge 
formed; during the second, the Galactic disk slowly grew, in the framework of 
an inside-out scenario of formation (Matteucci \& Fran\c cois 1989). 

We assume the Li content observed in stars belonging to the Spite plateau as 
indicative of the primordial value (A(Li)$_{\mathrm{p}}$ $\sim$ 2.2 dex; 
Bonifacio \& Molaro 1997) and justify the rise off the plateau as due to Li 
injection into the ISM from several lithium factories (C-stars, massive AGB 
stars, Type II SNe, nova systems). The possible contribution to $^7$Li 
production by cosmic rays is not taken into account.

We ran three models:
\begin{itemize}
\item model A, accounting for $^7$Li production from C-stars, massive AGB 
stars and Type II SNe (best model in Matteucci et al.\,1995, 1999);
\item model B, the same as model A with the addition of Li synthesis from 
thermonuclear runaway in nova outbursts (see Romano et al.\,1999);
\item model C, like model B but without any contribution from C-stars and with 
a lower Li production from both massive AGB stars and SNeII.
\end{itemize}
Model results are compared to the observational data in Fig.1.

\section{Model results and conclusions}

Let us first point out how using a single category of stellar lithium 
producers, we can not explain all the features of the observed diagram: for 
example, nova systems, evolving on long timescales ($\ge$ 1 Gyr), cannot 
account for a rise of the Spite plateau before a metallicity of about -\,0.5 
dex, but they are able to explain the highest abundances measured in the most 
metal-rich stars (see Fig.3). Opposite considerations hold for both AGB stars 
and Type II SNe, acting on shorter timescales but being able to explain only 
1/2 of the observed present ISM $^7$Li abundance.

\begin{figure}
\plotone{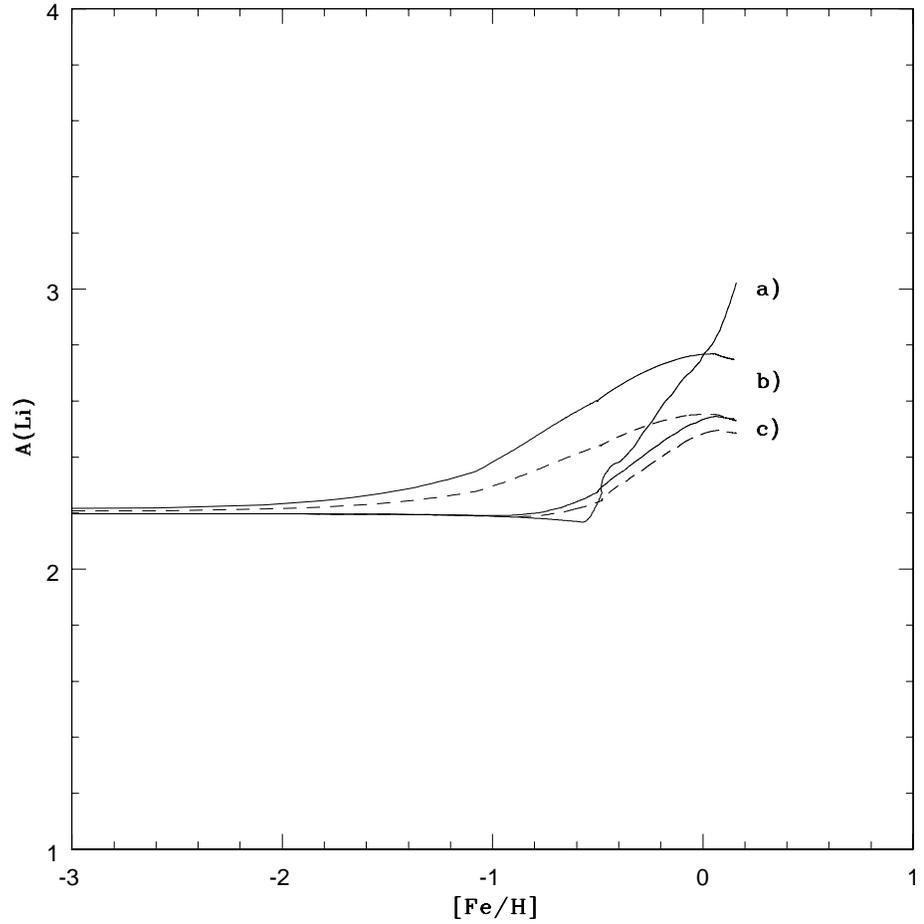}
\caption{ A(Li) vs [Fe/H] theoretical trends for the solar neighborhood as 
predicted when only a single category of stellar lithium producers is allowed: 
a) nova systems (yields from Jos\'e \& Hernanz 1998); b) Type II SNe (yields 
from Woosley \& Weaver 1995; straight line: total yields, dashed line: yields 
reduced to a half); c) AGB stars (prescriptions from Matteucci et al.\,1995; 
straight line: massive AGB stars plus C-stars, dashed line: only massive AGB 
stars).}
\end{figure}

\begin{figure}
\plotone{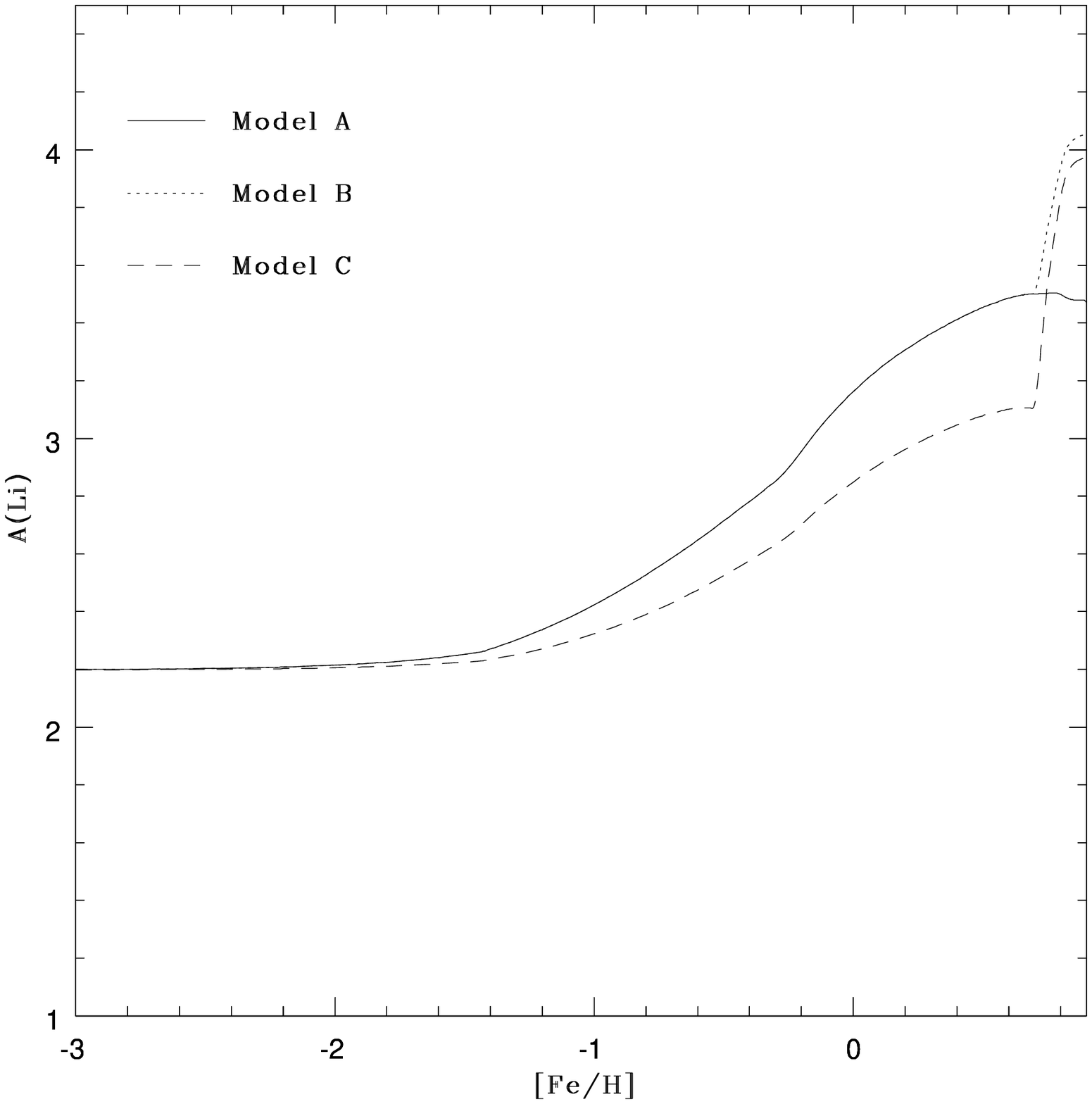}
\caption{ A(Li) vs [Fe/H] theoretical diagram for the bulge. Model B and C, 
taking into account $^7$Li production by novae, predict a steep rise at the 
highest metallicities. Similar trends are due to the interplay between the 
long timescales of Li production in novae and the short timescale of bulge 
formation.}
\end{figure}

So, we have tried to reproduce the upper envelope outlined by the experimental 
points by considering all the possible stellar sources together in the 
chemical evolution model (Fig.1). Model A, accounting for all the sources but 
novae, fails to reproduce the highest values observed in the more metal-rich 
dwarfs. To succed in doing this, we need to add nova systems (model B = model 
A + novae); in that case, also lowering the contribution by AGB stars and 
supernovae a satisfactory result can be found, the only difference with 
respect to the previous one being a slower rise off the plateau (model C).
Models accounting for nucleosynthesis from novae predict an increase in the 
ISM $^7$Li abundance from the epoch of the Solar System formation up to now. 
On the contrary, if we rule out novae as lithium producers, we find a flat 
trend. We suggest that the contribution to Li enrichment from neutrino-induced 
nucleosynthesis in Type II SN explosions is probably not so significant as 
assumed in previous chemical evolution models, although the paucity of data in 
the region -\,1.5 $\le$ [Fe/H] $\le$ -\,0.5 does not allow us to draw any firm 
conclusion. Our main conclusion is that a significant $^7$Li production from 
nova systems seems to be strongly requested to gain the high lithium 
abundances observed in the most metal-rich stars of the sample.

Finally, let us have a look at the situation predicted for the Galactic bulge. 
This central region of the Galaxy is thought to be the result of a faster 
collapse with a more efficient star formation, so that metallicities higher 
than the present one in the solar neighborhood should be gained in a shorter 
timescale. Therefore, when novae start restoring their newly synthesized 
$^7$Li, a metallicity as large as $\sim$ 0.5 dex is likely to have been already
 achieved in the gas in the bulge. A present Li abundance as high as 4 dex 
should be expected in the most metal-rich (and of course undepleted) bulge 
stars. If novae are suppressed, the present Li content in these stars should 
be 3.5 dex (see Fig.4). For testing similar predictions we need observations 
of the Li\,I feature in a statistical significant sample of very peculiar 
objects: metal-rich bulge stars which have never suffered lithium depletion. 
This is a hardly achievable goal, because we know that stars become more and 
more efficient in destroying their initial lithium content already in the 
pre-main sequence phase as long as their metallicity increases (e.g.\,Ventura 
et al.\,1998). The recent detection of the Li\,I line in the bulge star 
97\,BLG\,45 by Minniti et al.\,(1998) is at the moment the only one of this 
type, and unfortunately shows a star which is likely to have depleted its 
initial lithium content (Matteucci et al.\,1999). So, observational efforts in 
this sense will be very welcome.

\acknowledgements I would like to thank F.\,Matteucci, P.\,Molaro, 
P.\,Bonifacio, F.\,D'Antona and J.\,Danziger for valuable contributions 
in the course of this work.

\end{document}